\documentclass[twocolumn,secnumroman,floatfix,amssymb,amsmath,nofootinbib,tightenlines,showpacs,superscriptaddress,aps,prb]{revtex4}
\usepackage{amsfonts}
\usepackage[colorlinks=true,linkcolor=blue,pagecolor=blue,filecolor=blue,menucolor=yellow,urlcolor=blue,citecolor=blue,anchorcolor=blue]{hyperref}%
\usepackage{graphicx}
\usepackage{dcolumn}
\usepackage{times}
\usepackage{soul}
\usepackage{sidecap}

\newcommand{\bq}{\begin{equation}}
\newcommand{\eq}{\end{equation}}

\newcommand{\etal}{\emph{et al.}}
\newcommand{\pr}{Phys. Rev. }

\begin{document}

\title{Singlet-Triplet Superconducting Quantum Magnetometer}

\author{Mohammad Alidoust }
\email{phymalidoust@gmail.com} \affiliation{Department of Physics,
Norwegian University of Science and Technology, N-7491 Trondheim,
Norway}

\author{Klaus Halterman}
\email{klaus.halterman@navy.mil} \affiliation{Michelson Lab, Physics
Division, Naval Air Warfare Center, China Lake, California 93555}

\author{Jacob Linder}
\email{jacob.linder@ntnu.no} \affiliation{Department of Physics,
Norwegian University of Science and Technology, N-7491 Trondheim,
Norway}
\date{\today}

\begin{abstract}
Motivated by the recent experimental realization of a quantum
interference transistor based on the superconducting proximity
effect, we here demonstrate that
the inclusion of a textured ferromagnet both strongly enhances the
flux sensitivity of such a device and additionally allows for
singlet-triplet switching by tuning a bias voltage. This
functionality makes explicit use of the induced spin-triplet
correlations due to the magnetic texture. Whereas the existence of
such triplet correlations is well-known, our finding demonstrates
how spin-triplet superconductivity may be utilized for concrete
technology, namely to improve the functionality of ultra-sensitive
magnetometers.
\end{abstract}


\maketitle

\section{Introduction}

The synthesis of magnetism and
superconductivity (SC) is intriguing both on a fundamental
level and in terms of possible new devices based on the interplay
between these phenomena. The former aspect encompasses a wide
range of areas in condensed matter physics ranging from
heavy-fermion compounds with several coexisting long-range orders
\cite{cite:saxena_nature_00} to unusual forms of
SC emerging in hybrid structures,
displaying a resilience towards pair-breaking from impurities
and 
paramagnetic limitations \cite{cite:berizinskii, 
cite:bergeret_prl_01}. On the other hand, the Superconducting
Quantum Interference Device \cite{cite:squid} is a prominent 
example of how the interaction between magnetism and
SC may give rise to important
functionality in technology
\cite{cite:giazotto2,cite:giazotto1,cite:Spathis}.

In a broader context, the emergence of multiple spontaneously broken
symmetries is pivotal since it lies at the heart of a variety of
phenomena outside the field of condensed matter. By creating
heterostructures of ferromagnets (FM) and superconductors (S), such a situation is obtained due to the proximity
effect \cite{cite:degennes,cite:mcmillan}: the induction of
superconducting correlations in a magnetic material and vice versa
\cite{cite:buzdin_rmp_05,
cite:bergeret_rmp_05,cite:klaus2,cite:zhou_hammer}. One particular
manifestation of the proximity effect emerges in an inhomogeneous
FM which generates spin-polarized
Cooper pairs with a large penetration depth despite the exchange
field \cite{cite:bergeret_rmp_05, cite:robinson}. Whereas the
existence of such spin-triplet superconducting correlations is
well-known, it has remained unclear if they may be utilized for
functional purposes in technology.

In this paper, we show that such proximity-induced spin-triplet
correlations can be utilized to obtain an ultra-sensitive
interferometer with a highly stable flux-sensitivity that
utilizes singlet and triplet
proximity effects, both controlled via a bias voltage. We will refer to this 
device as a Singlet-Triplet Superconducting Quantum Magnetometer
(SQM). The most important aspect is that it greatly enhances the
range over which there is a flux-sensitivity compared to previously
experimentally realized magnetometers
\cite{cite:giazotto1,cite:giazotto2}. One merit of this finding is
that it demonstrates how spin-triplet
SC may be utilized for a concrete
technology, namely to improve the functionality of ultra-sensitive
magnetometers which could lead to advancements in low temperature nanoscale spintronics that require minute flux detection. A similar interferometer was experimentally
realized in
\cite{cite:giazotto1,cite:giazotto2,cite:giazotto3,cite:meschke},
albeit \textit{without} any ferromagnetic element. Incorporating a textured ferromagnet in this type of device should therefore be feasible,
and hence would allow for a test of our predictions. In addition, the considerable improvement in sensitivity in turn presents new possibilities for many practical
applications spanning numerous disciplines, including
medicine (neuromagnetic studies of the brain), geophysics (where field mapping is needed with high precision), and components of
magnetoresistive devices (including memory cells).

We will demonstrate that the
inclusion
of an inhomogeneous FM having a
spatially dependent magnetic texture, considerably improves the
functionality of such an interferometer. In fact, the
SQM features a stable phase-sensitivity which
should be contrasted with the scenario without any ferromagnetic
elements considered previously, in which case the sensitivity is
finite only for special values of the external flux. As we will
explain in detail, this effect originates from the triplet
correlations present in the junction. To make
our discussion as general as possible, we cover both the diffusive
and ballistic regimes of transport, utilizing the Keldysh-Usadel
\cite{cite:usadel} and Bogoliubov-de Gennes (BdG)
\cite{cite:degennes} formalisms.

\begin{figure}[t!]
\includegraphics[width=8.5cm,height=5.80cm]{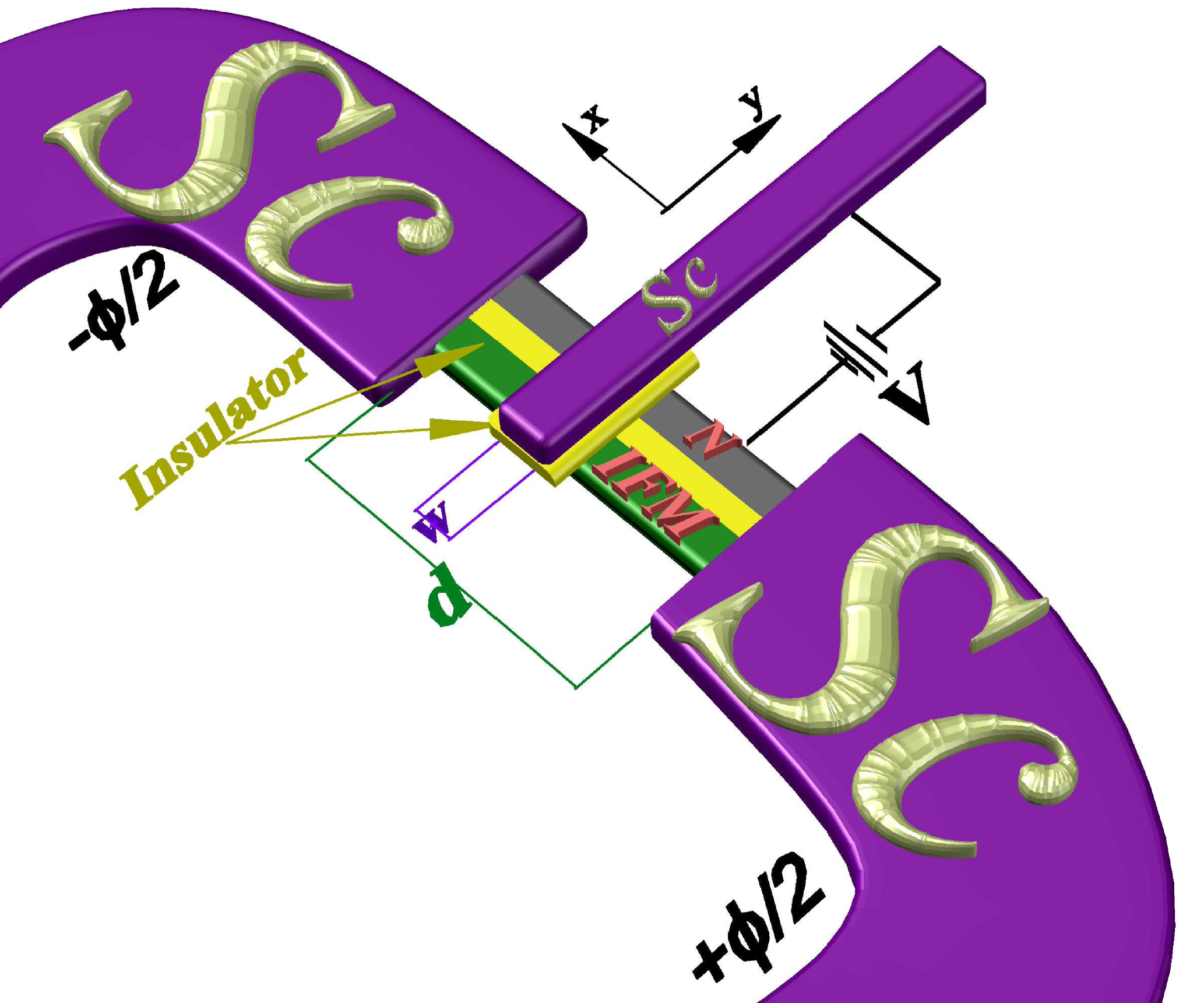}
\caption{\label{fig:model} Schematic set up of the proposed
SQM. An inhomogeneous FM and a
normal-metal (N) nanowire each of length $d$ are connected in
parallel with an insulating layer separating them. This structure is
sandwiched between two bulk $s$-wave superconductors
along the $x$-axis. A third S electrode of width $W$
is deposited on top of the nano-wires along the $y$-direction. By
applying a voltage $V$, a quasiparticle current flows between the
electrode and nano-wires (we refer to the third
superconducting electrode as the ``collector electrode"). The bulk 
S are assumed to be part of a closed
loop circuit threaded by an external flux, so that any supercurrent
flowing into the collector electrode can be neglected. The external
flux induces a macroscopic phase difference, $\phi$, between the
bulk S.}
\end{figure}

\section{Theory}

Assuming an applied voltage difference, $V$, between the
non-superconducting region and the collector electrode (see
Fig.~\ref{fig:model}), and neglecting Coulomb blockade
effects, the resistive quasiparticle current is given by the
following relation \cite{cite:likharev,cite:giazotto3,cite:vasenko}:
\begin{align}
eRI_{qp} &= \frac{1}{W}\int_{-W/2}^{+W/2}\int_{-\infty}^{\infty}dx
d\varepsilon N(x,\varepsilon,\phi,T)N_S(\varepsilon-eV)\notag\\
&\times(f(\varepsilon-eV)-f(\varepsilon)),
\end{align}
where $N(x,\varepsilon,\phi,T)$ is the local density of states (DOS)
in the sandwiched region normalized by its value in the normal state
and $N_S$ is the collector electrode's DOS which can be expressed by
\begin{align}
N_S(\varepsilon)=\text{Re}\left\{|\varepsilon|/\sqrt{\varepsilon^2-\Delta(T)^2}\right\}
\end{align}
in the $s$-wave case. Here,
$f(\varepsilon)$ is the
Fermi-Dirac energy distribution, 
$\varepsilon$ is quasiparicle energy measured
from the Fermi level, $\phi$ is the superconducting phase difference
between the bulk superconducting leads, $e$ is the unit electronic
charge, and $R$ is the resistance of the non-superconducting
nano-wire. The integration is taken over the collector electrode
width in the $x$-direction 
(see Fig.~\ref{fig:model}). We note in passing that if a voltage difference is present between the S banks,
it is possible to induce Shapiro steps even in the absence of external radiation \cite{cite:cuevas}. 
In our setup, however, 
no voltage shift exists between the bulk S regions.

To be concrete, we consider a conical texture for the FM and adopt a
model relevant for Ho where the magnetic moment rotates on the
surface of a cone with apex angle $\alpha$$=$$4\pi/9$ and turning
angle $\varpi$$=$$\pi/6$. If we assume that the distance of
interatomic layers is $a$$=$$0.02d_F$ \cite{cite:alidoust1},
the exchange field, $\textbf{h}$, 
can be written as
$\textbf{h}$$=$$h(\cos\alpha\hat{x}+\sin\alpha[\sin(\varpi
x/a)\hat{y}+\cos(\varpi x/a)\hat{z}]).$
In the diffusive regime, the mean free path is much smaller
than junction length \cite{cite:usadel}. In what follows, we
consider the full proximity effect of the diffusive regime without
making any simplifying assumptions such as low interface
transparency or linearization of the equations. We emphasize that
our BdG approach leads to the same results in the ballistic limit.
The Usadel equation \cite{cite:usadel} can be compactly written
as
\begin{eqnarray}
D[\hat{\partial},\hat{G}[\hat{\partial},\hat{G}]]+i[ \varepsilon
\hat{\rho}_{3}+
\text{diag}[\textbf{h}\cdot\underline{\sigma},(\textbf{h}\cdot\underline{\sigma})^{\tau}],\hat{G}]=0.
\end{eqnarray}
Here $D$ is the diffusion constant, $\hat{G}$ represents the total
Green's function and $\hat{\rho}_{3}$ and $\underline{\sigma}$ are
$4$$\times$$4$ and $2$$\times$$2$ Pauli matrixes, respectively. The
Usadel equation
is supplemented by the Kupriyanov-Lukichev 
boundary conditions at interfaces along the $x$-axis
\cite{cite:zaitsev};
\begin{align}
2\zeta\hat{G}\hat{\partial}\hat{G}=[\hat{G}_{\text{BCS}}(\phi),\hat{G}],
\end{align}
in which $\hat{G}_{\text{BCS}}$ is the bulk solution and $\zeta$
controls interface opacity. To have more stability in numerical
solutions, we use the so-called Ricatti parametrization of Green's
function where the local DOS can be expressed by \cite{cite:zhou_hammer,cite:alidoust2}:
\begin{align}
N(x,\varepsilon,\phi,T)=\frac{1}{2}\text{Re}\left\{\text{Tr}[(1-\gamma^{R}\tilde{\gamma}^{R})/(1+\gamma^{R}\tilde{\gamma}^{R})]\right\}.
\end{align}
To model realistic interfaces, we consider a finite barrier transparency by setting $\zeta=4$: a perfect interface has $\zeta=0$ whereas the tunneling limit is reached for $\zeta\gg1$. The value of $\zeta$ thus controls
the magnitude of the proximity effect, but it has no bearing on our main result. We use $\hbar$$=$$k_B$$=$$1$ and set a fixed temperature
throughout our computations equal to $0.05T_c$. All lengths in the
system are normalized by $\xi_S$, the superconducting
coherence length.

\section{Results and Discussion}

\begin{figure*}
 \centering
\includegraphics[width=18.5cm,height=8.8cm]%
{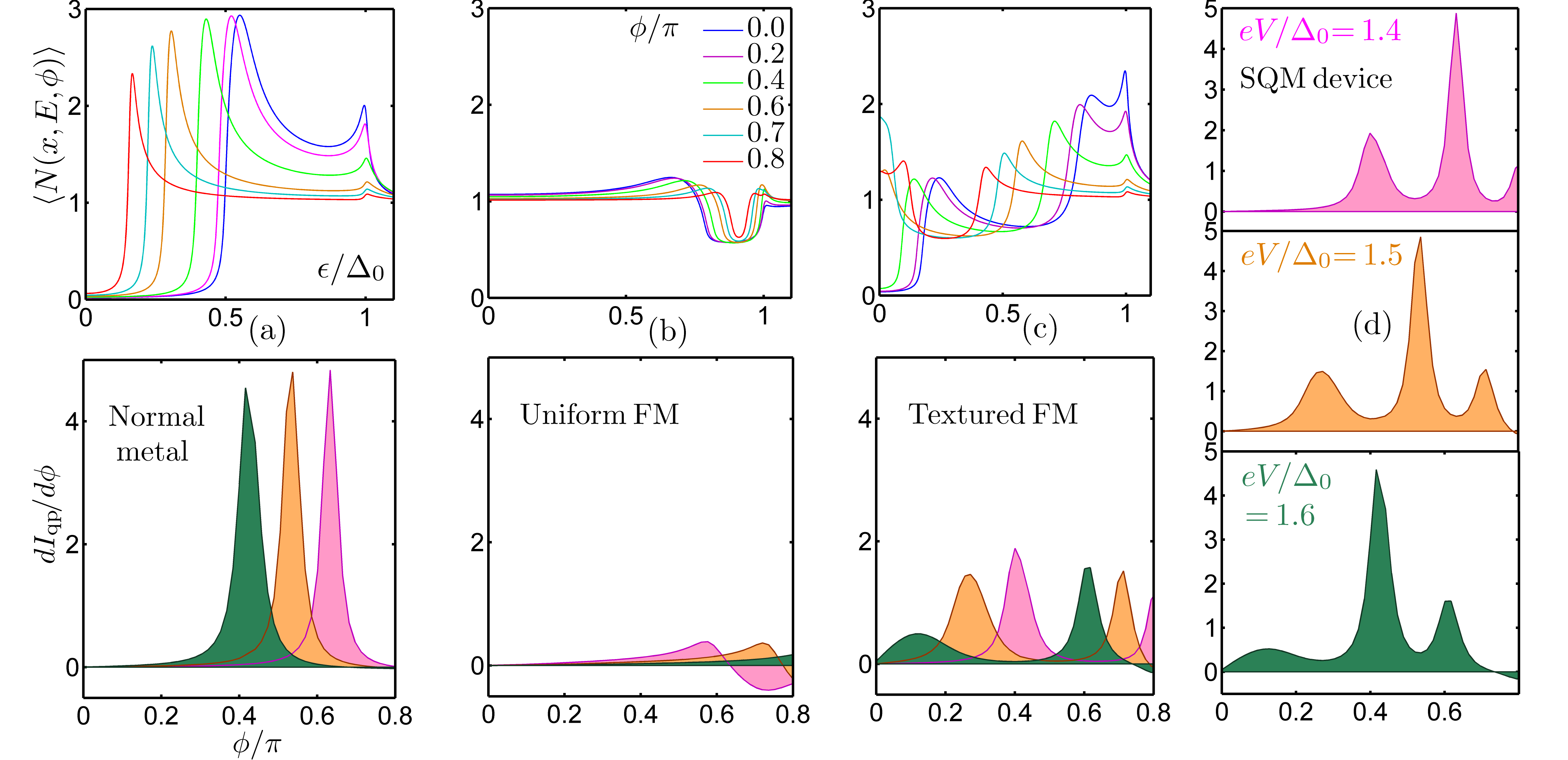} \caption{(Color online)  Top row (a)-(c): DOS vs. quasiparticle energy $\epsilon$ for several
phase differences $\phi$.
The columns correspond to (a) a S/N/S junction,
(b) a S/uniform FM/S junction, and (c) a S/conical FM/S junction.
The DOS has been averaged over the collector electrode width $W=d/2$
and $d=0.5\xi_S$.
Bottom row (a)-(c): the device sensitivity $dI_{qp}/d\phi$ vs. phase difference $\phi$ for different choices of bias voltage $eV/\Delta_0=\{1.4,1.5,1.6\}$ with $d=0.3\xi_S$.
In the column (d), we show the sensitivity for our proposed SQM device which incorporates both a N and conical FM nano-wire (see Fig. \ref{fig:model}).} \label{fig:sensitivity}
\end{figure*}

Since the underlying physical mechanism for the magnetometer is the modulation of the DOS in the nano-wire when changing the superconducting phase difference, we first show the results for this quantity in the top row of Fig. \ref{fig:sensitivity}. The DOS is averaged over the width of the collector electrode and plotted vs. quasiparticle energy. The magnitude of the exchange splitting is set to
$h$$=$$3\Delta_0$ whereas the length of the nano-wires is set to $d$$=$$0.5\xi_S$, 
which typically corresponds to $\sim 10$ nm. The columns correspond to (a) a S/N/S junction, (b) a S/uniform FM/S junction, and (c) a S/conical FM/S junction. For the S/N/S junction in (a), we obtain the
well-known minigap profile in the DOS spectrum
\cite{cite:zhou_hammer,cite:giazotto3} whereas the minigap vanishes
in a uniform FM junction shown in (b). This changes qualitatively
for the conical FM junction in (c): the DOS now peaks near the Fermi level for
some phase differences. This resonant behavior of the DOS in the
inhomogeneous case is a manifestation of the presence of 
spin-triplet superconducting correlations which in the diffusive
limit have an odd-frequency symmetry
\cite{cite:bergeret_rmp_05,cite:kontos_prl_01}.

To see how this influences the flux sensitivity, consider now the bottom rows of Fig.~\ref{fig:sensitivity}.
Columns (a)-(c) correspond to the same type of junction as described above, and we have plotted the device sensitivity $dI_{qp}/d\phi$ as a 
function of the phase difference with $d$$=$$0.3\xi_S$. The flux sensitivity 
can be directly related to the DOS shown in the top row of Fig.~\ref{fig:sensitivity}. To see this, consider first the normal and uniform FM cases in (a) and (b). The magnitude of $dI_{qp}/d\phi$ is
highly suppressed for the uniform magnetization
case for all
values of voltage difference $eV/\Delta_0$ compared to normal
junction. The superconducting correlations are
effectively washed out due to destructive competition between 
singlet and short-range triplet correlations which 
suppresses the proximity effect overall, leading to a featureless DOS and resulting low device sensitivity
\cite{cite:alidoust2}.

For the normal metal case, however, the
flux-sensitivity is seen to feature a peak at a fixed phase. It should be noted that a \textit{single} peak always appears completely irrespective of the parameters chosen for the normal metal case. This
means that whereas the sensitivity is high, it is restricted to a
very narrow regime of external flux so that it would be insensitive
to any magnetic fluxes deviating from this special value.

Consider now instead a junction with a conical FM shown in 
Fig.~\ref{fig:sensitivity}(c), where long-range spin-triplet correlations are present. The sensitivity 
$dI_{qp}/d\phi$ now shows a strongly enhanced stability with respect to
$\phi$, which is present for several voltages.
This can be clearly seen by noting the presence of two peaks in $dI_{qp}/d\phi$, which implies an improved device sensitivity by $\sim100\%$ compared to (a). Although the sensitivity amplitude is now somewhat smaller than the normal junction, the
conical FM renders the magnetometer to have a flux-sensitivity covering a
large part of the flux-regime. We will show below that this 
property used in conjunction with the singlet proximity effect from
a normal metal produces a magnetometer with sensitivity for
essentially \textit{any} flux.

We first explain the physical origin of this enhanced
flux-sensitivity arising from the presence of a textured
FM. It is known that magnetic
inhomogeneities 
can contribute to proximity induced triplet pairing
in superconducting structures
\cite{cite:bergeret_rmp_05}. When this occurs, the DOS differs
qualitatively from the non-magnetic superconducting proximity
effect. As seen in the top panel of Fig.~\ref{fig:sensitivity}(a), the the energy-resolved DOS for the 
non-magnetic case
simply exhibits a
minigap, while in (c)
a much stronger
variation is observed. In particular, when the
triplet correlations are present, a
zero-energy peak arises rather than a gap as one tunes the
superconducting phase difference $\phi$. This can be traced back to
the odd-frequency symmetry of the triplet correlations, which
enhances the low-energy spectral weight. It is clear from the DOS of
the textured FM case why the quasiparticle
current is more sensitive to 
$\phi$ than in the previous N
and uniform F cases: changing 
 $\phi$
alters the DOS not only by simply closing a minigap, as in the N 
case, but by inducing a transition from a fully gapped to a peaked
DOS at zero energy in addition to altering the spectral features at
other subgap energies. 
By ``gapped" we mean an absence of single-particle states in the low-energy spectrum. 
 The net result is a magnetometer sensitivity $dI_{qp}/d\phi$ which
takes on appreciable values over a large interval of fluxes rather
than only peaking near a specific value of the flux. 
These type of spin-polarized superconducting correlations in the 
system thus improves the device functionality.

Having now described how the presence of textured ferromagnetism
alters the magnetometer sensitivity, we proceed to introduce the
SQM device. Consider a situation where a N and
a conical FM wire are connected in parallel as shown in Fig.~\ref{fig:model}. An insulating layer prevents
interference between the two wires.
When the connected wires are in electrical contact with a top
superconducting electrode, a resistive quasiparticle current
flows between them in the presence of an applied bias voltage $V$.
We underline here that this resistive current is naturally separated
from the supercurrent flowing in the superconducting loop due to
current conservation. Let us denote the DOS of the N and conical FM regions
by $N_{\text{N}}$ and $N_{\text{FM}}$. The energy-resolved
quasiparticle current flowing to the top superconducting electrode
will then be proportional to $N_{\text{N}}$$+$$N_{\text{FM}}$. The
total current is obtained by integrating over all energies and
taking into account the proper Fermi-Dirac distribution functions.
The resultant 
flux-sensitivity, $dI_{qp}/d\phi$, of the proposed
SQM is shown in Fig. \ref{fig:sensitivity}(d).

Remarkably, the sensitivity is
finite over almost the entire range of phase-differences, which
means that the SQM would operate as a magnetometer
over a much broader flux-range than in the experiments of Refs.
\onlinecite{cite:giazotto1,cite:giazotto2,cite:giazotto3}. This advantage
stems from the fact that the sensitivity $dI_{qp}/d\phi$ has its
main contribution from \textit{different} values of the phase (flux)
in the N and conical FM cases, which is generated by the triplet 
correlations: the peaks in Fig. \ref{fig:sensitivity}(a) and (c) occur at incommensurate values and therefore their combination yields a total
sensitivity which covers essentially the entire range of fluxes
and leads to a $\sim 200\%$ improvement of the
device sensitivity. This idea can be extended to a device
with several layers where an incommensurate sensitivity of each layer would
generate a huge enhancement of the device efficiency. Another interesting aspect of the proposed device is that for longer junction lengths, 
the contribution to the sensitivity $dI_{qp}/d\phi$ from the singlet and triplet proximity effects occurs in different voltage regimes. 
For voltages close to $eV \sim \Delta$, the singlet proximity effect dominates whereas for larger voltages $eV > 1.5\Delta$,
the triplet proximity effect is responsible for the device sensitivity. This means
that the contribution from the N and conical FM  region
can be individually separated by tuning the voltage $eV$,
effectively switching the two wires ``\textit{on}" and
``\textit{off}". 

We now turn to the ballistic regime of transport and utilize the BdG
technique which enables us to fully isolate all triplet  pairing
correlations in the system and investigate the precise behavior of the proximity-induced
triplet superconducting correlations, both for equal- and opposite
spin-pairing. In terms of the quasiparticle amplitudes
$u_n^\sigma$ and $v_n^\sigma$ with excitation energy
$\varepsilon_n$, the BdG equations are written
\cite{cite:degennes,cite:klaus1}:
\begin{equation}\label{BdG}
\left(
  \begin{array}{cc}
    \mathcal{H}_M-\mu\mathbb{I} & \Delta(x)\mathbb{I}\\
    \Delta^{\ast}(x) \mathbb{I} & -(\mathcal{T}\mathcal{H}_M\mathcal{T}^{-1}-\mu\mathbb{I}) \\
  \end{array}
\right) \Psi_n =\varepsilon_n \Psi_n,
\end{equation}
where the wavefunction reads
\begin{align}
\Psi_n\equiv
(u_n^\uparrow,u_n^\downarrow,v_n^\uparrow,v_n^\downarrow)^T,
\end{align}
whereas the normal-state Hamiltonian is defined by:
\begin{align}
{\mathcal
H}_M=-\vec{\nabla}^2_{x}/2m+\varepsilon_\perp-{\bf
h} \cdot \underline{\sigma}.
\end{align}
Above, $\mu$ is the chemical potential,
and $\varepsilon_\perp$ represents the transverse kinetic energy.
The time reversal operator, $\mathcal{T}$, is written ${\mathcal
T}$$=$$i \sigma_z {\mathcal C}$, where ${\mathcal C}$ is the
operator of complex conjugation. We take the pair potential,
$\Delta(x)$, to be piecewise constant in the S regions with a phase
difference $\phi$. To capture the triplet correlations, we align the
quantization axis with the local magnetization vector in the
conical FM.

\begin{figure}[t!]
\includegraphics[width=9.0cm,height=3.6cm]{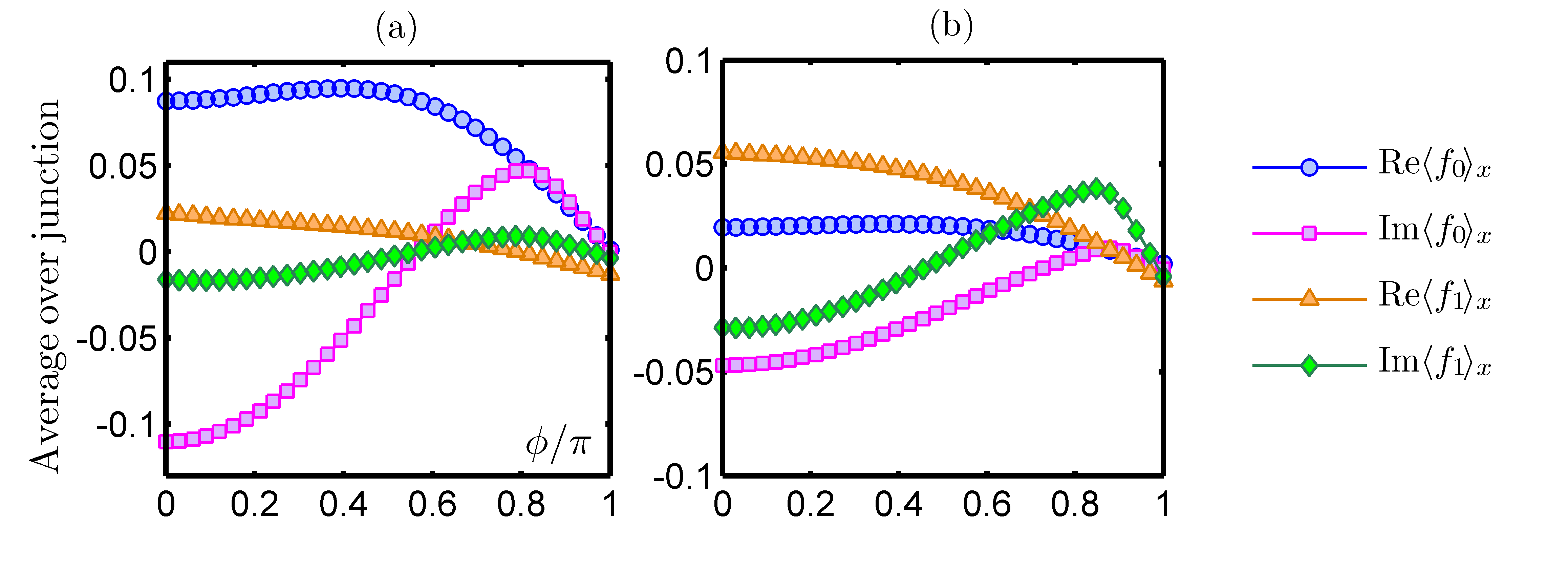}
\caption{\label{fig:mixed} Imaginary and real parts of spin-triplet
superconducting correlations averaged over the junction length
plotted vs $\phi$ in the ballistic regime where a conical FM is
sandwiched between two SC banks. (a)
$d$$=$$0.3\xi_S$ and (b) $d$$=$$0.5\xi_S$.}
\end{figure}

The time-dependent
triplet correlations with opposite ($f_0$) and equal ($f_1$)
spin-pairings, are then expressed as:
\begin{align}
f_0 &= \frac{1}{2}\sum_n\bigl\{
\cos\alpha f_n^{+} +\sin \alpha [\cos({\varpi x}/{a}) (f_n^{\uparrow\downarrow}-f_n^{\downarrow \uparrow}) \notag\\
&\hspace{1.2cm}+ i \sin({\varpi x}/{a}) f_{n}^{-} ]\bigr\}\zeta_n(t),\notag\\
f_1 &= \frac{1}{2}\sum_n\bigl\{
\sin\alpha f_n^{+} -\cos \alpha [\cos({\varpi x}/{a}) (f_n^{\uparrow\downarrow}-f_n^{\downarrow \uparrow}) \notag \\
&\hspace{1.2cm}+ i \sin({\varpi x}/{a}) f_{n}^{-} ]\bigr\}\zeta_n(t),
\end{align}
in which we define
\begin{align}
f_{n}^{\pm} &\equiv f_n^{\uparrow
\uparrow}\pm f_n^{\downarrow \downarrow}; \qquad
f_n^{\sigma \sigma'} \equiv u_n^{\sigma}
v_n^{\sigma'\hspace{-.06cm}*}.
\end{align}
We also have introduced the time-dependent quantity \cite{cite:klaus2}
\begin{align}
\zeta_n(t)=\cos({\varepsilon_nt})-i\sin({\varepsilon_nt})
\tanh(\varepsilon_n/(2 T)).
\end{align}
When $t$$=$$0$, the
triplet correlations vanish by virtue of the Pauli exclusion
principle. As $t$ is increased, the triplet correlations generally
form near the F/S interfaces, and slowly grow in amplitude. In what
follows, we scale $t$ by a characteristic ``Debye'' energy,
$\omega_D$, and set $\omega_D t$$=$$8.8$ as a representative choice
which allows one to witness the extended behavior of the triplets in
the junction. When the junction contains a conical FM, the equal 
spin-pairing components are generally non-zero and their evolution
against phase difference is shown in Fig.~\ref{fig:mixed}. For
$d$$=$$0.3\xi_S$ (left panel), it is seen that the averaged opposite
spin-pairing components are larger in magnitude and have the
greatest variation with phase, whereas both types of triplet
correlations feature a non-monotonic behavior vs $\phi$, nearly
vanishing at $\phi$$=$$\pi$. As we increase the junction length to 
$d$$=$$0.5\xi_S$ however, it is seen that the $f_1$ correlations are
now more prominent. This is consistent with the fact that the
equal-spin correlations have a long-ranged penetration depth in
spite of the presence of an exchange field, whereas the
opposite-spin triplet pairs have increased oscillations that decay
over a length scale similar to that of the spin-singlet state,
leading to reduced spatial averages.

\section{Summary}

In summary, we have shown that long-ranged spin-triplet superconducting correlations can be utilized for practical 
technologies and offer advancements
in the form of improved sensitivity for magnetometers.
The key element is the inclusion of a textured ferromagnet
and a
normal-metal which alters the superconducting proximity
effect and allows for a $\sim 200\%$ increased flux-operation range for the proposed device.
We demonstrate a highly stable flux-sensitivity that
utilizes both singlet and triplet
proximity effects. We
found a considerable enhancement of the
range over which there is a flux-sensitivity compared to previously
fabricated magnetometers in which there are no ferromagnet elements.

\acknowledgments
K.H. is supported in part by IARPA
and by a grant
of supercomputer resources provided by the DOD HPCMP. J.L. was supported by the COST Action MP-1201 "Novel Functionalities through Optimized Confinement of Condensate and Fields".

\end{document}